\documentclass[5p]{elsarticle}
\usepackage{graphicx} 
\usepackage{hyperref}
\usepackage{natbib}
\usepackage{geometry}
\usepackage{float}
\usepackage{chngcntr}
\usepackage{amsmath,amsfonts,amssymb,amsthm}   
\usepackage{algorithm}
\usepackage{algorithmic}

\usepackage{booktabs}
\usepackage{xfrac}
\usepackage{tabularx}
\usepackage{lipsum}
\usepackage{epstopdf}
\usepackage{algorithmic}
\usepackage{bm}
\usepackage{upgreek}
\usepackage{pstricks}
\usepackage{enumitem}
\usepackage{bbold}
\newlist{steps}{enumerate}{1}

\title{Large scale response of a vehicle wake to on-road perturbations}
\date{April 2024}

\author[1,3]{Agostino Cembalo}
\ead{agostino.cembalo@stellantis.com}
\author[1]{Jacques Borée\corref{cor1}}
\ead{jacques.boree@ensma.fr}
\author[2]{Patrick Coirault}
\ead{patrick.coirault@univ-poitiers.fr}
\author[3]{Clément Dumand}
\ead{clement.dumand@stellantis.com}
\cortext[cor1]{Corresponding author}

\affiliation[1]{organization={Pprime Institute CNRS, ENSMA, University of Poitiers},
addressline={1 Av. Clément Ader},
postcode={86360},
city={Chasseneuil-du-Poitou},
country={France}}
\affiliation[2]{organization={Laboratory LIAS - ENSIP, University of Poitiers},
addressline={2 rue Pierre Brousse},
postcode={86073},
city={Poitiers},
country={France}}
\affiliation[3]{organization={STELLANTIS, Advanced Innovation},
addressline={212 Bd Pelletier},
city={Carrières-sous-Poissy},
postcode={78955},
country={France}}

\begin{document}

\theoremstyle{plain}
\setcounter{assumption}{0}

\begin{abstract}
The aim of this research work is to analyse the large scale response of a vehicle wake to on-road perturbations by using an instrumented vehicle and a combination of scale one wind tunnel tests, track trials and on road experiments. More precisely, in all these tests, we focus on the analysis of the asymmetry of the pressure distribution at the base. Proper Orthogonal Decomposition (POD) is used. For all cases considered, POD analysis reveals two dominant modes, respectively associated with vertical and horizontal wake large scale reorganisation. More than 50\% of the total energy is carried by these two modes and this value increases significantly for on-road tests. Noteworthy, the low-frequency energy content of the temporal coefficients of these modes is significantly higher on-road. Low frequencies (even very low ones) then play a major role, corresponding to a quasi-static perturbation domain of the velocity seen by the vehicle. We show that a quasi-steady exploration of the on-road yaw angle statistical distribution during a wind tunnel test captures phenomena similar to those observed on the road and is therefore interesting to evaluate on-road aerodynamic performances. This also opens perspectives for developing closed loop control strategies aiming to maintain a prescribed wake balance in order to reduce drag experienced on the road.

\end{abstract}

\maketitle

\section{Introduction}
\label{sec:introduction}

The European Environment Agency has assessed the impact of transports on greenhouse gases and state that 
"\emph{Transport is responsible for a quarter of the EU's greenhouse
gas emissions, with road transport representing the greatest
share (72\% in 2019).}"
\footnote{https://www.eea.europa.eu//publications/transport-and-environment-report-2021\label{fn1}} Moreover, it is estimated that passenger cars are still the dominant transportation mean with a share of  approximately 80\% \footnote{https://www.eea.europa.eu/publications/transport-and-environment-report-2022/transport-and-environment-report/view\label{fn2}}. These studies highlight the importance of reducing the greenhouse gases emissions of road transportation and in this regards vehicle aerodynamics plays a crucial role. In fact, at highway speeds, approximately 70\% of the energy losses can be attributed to aerodynamic forces (\citet{Kadijk2012,Hucho1993}). In that respect, vehicle manufacturers put a lot of effort in optimizing vehicle's aerodynamics both with numerical simulations and wind tunnel testing. However, the vehicle's surroundings are constantly changing in real-life scenarios and the drag coefficient, usually optimised at zero yaw condition, varies continuously. 

As yaw angle is concerned, several studies have assessed an on-road average yaw angle distribution (\citet{Carlino2007, Cruz2017, Yamashita2017, Stoll2018}), showing that, on average, this distribution corresponds to a quasi-normal distribution centered around 0° with a standard deviation varying upon the external perturbations of the velocity seen by the vehicle due not only to wind turbulence but also to varying road surroundings such as guardrails, vehicles, bushes and trees among others. Hence, only considering the zero yaw angle drag coefficient translates into an under-prediction of the actual averaged drag coefficient. \citet{Howell2017} have addressed this problem by defining a wind averaged drag coefficient, on a WLTP (Worldwide Harmonised Light vehicles Test Procedure) road cycle, using quasi-steady approaches :
$$ C_{DWC} = 0.530 {C_D}_0 + 0.345{C_D}_5 + 0.130{C_D}_{10} + 0.007{C_D}_{15},$$ 
\noindent ${C_D}_i $ ($i = 0, 5, 10, 15^\circ$) being the drag coefficient at a given yaw angle and $C_{DWC}$ being the averaged drag coefficient based on a typical wind distribution. The authors of this study also specifically highlight the importance of reducing the sensitivity of the aerodynamic loads to the on-road perturbations. 

\citet{Cooper2007} and \citet{Watkins2007} successfully measured the unsteady flow characteristics seen by a road vehicle. They showed that the mean turbulence intensity is of the order of 5\% with integral length scales of the order of tens of meters. Moreover, turbulence seen by the vehicle has a rich spectra over a wide range of frequencies and it has been found that the wind gusts feed the spectrum in the low frequencies. Similar results have been found by \citet{Wordley2008}, \citet{Wordley2009} and \citet{Schrock2007}. The first two studies focus on the effect of four different terrain variations on the turbulence scales and intensities. It is shown that "\emph{the four different terrain overlap in the region of lower turbulence intensity  ($< 5\%$) and smaller turbulent length scale ($< 5 \;m$)}". The study of (\citet{Schrock2007}) emphasizes that, when the wind intensity grows, the energy content of the turbulence spectrum seen by the vehicle shifts to the low frequency region, resulting in about 75\% of the energy content at frequencies below $2 \;Hz$.

All these studies show the importance of on-road turbulence on vehicles aerodynamics. As an important follow-up on all these contributions, our objective in this paper is to provide data and to analyze the spatio-temporal distribution of the static pressure at the base of a vehicle submitted to such on road perturbations. In particular, it is interesting to determine if the external perturbations seen by the vehicle result in vertical or horizontal large scale asymmetries of the wake that we expect to contribute significantly to the mean drag experienced by the vehicle. To achieve this objective, we equipped a vehicle with a Prandtl antenna and a Conrad probe at the front as well as 49 pressure taps at the base. The vehicle was tested in a wind tunnel, on a track and a "RDE-like"  route (RDE: Real Driving Emissions). Through the collected data, we analyzed the pressure asymmetry of the wake and compared results between wind tunnel experiments and real-world road testing.

The paper is organised as follows: Section 2 gives information on the measurement systems installed in the car and Section 3 describes the different experimental setup. In Section 4 we focus on the wind tunnel experiments at zero yaw angle. Section 5 will highlight the major results obtained on-road and some comparison with the wind tunnel results will be made. Section 6 provides a statistical comparison between the on-road tests and a test with a dynamically varying yaw angle in the wind tunnel. Concluding remarks are proposed in Section 7.

\section{Measurement systems}
\begin{figure}[ht!]
    \centering
    \includegraphics[width=0.44\textwidth]{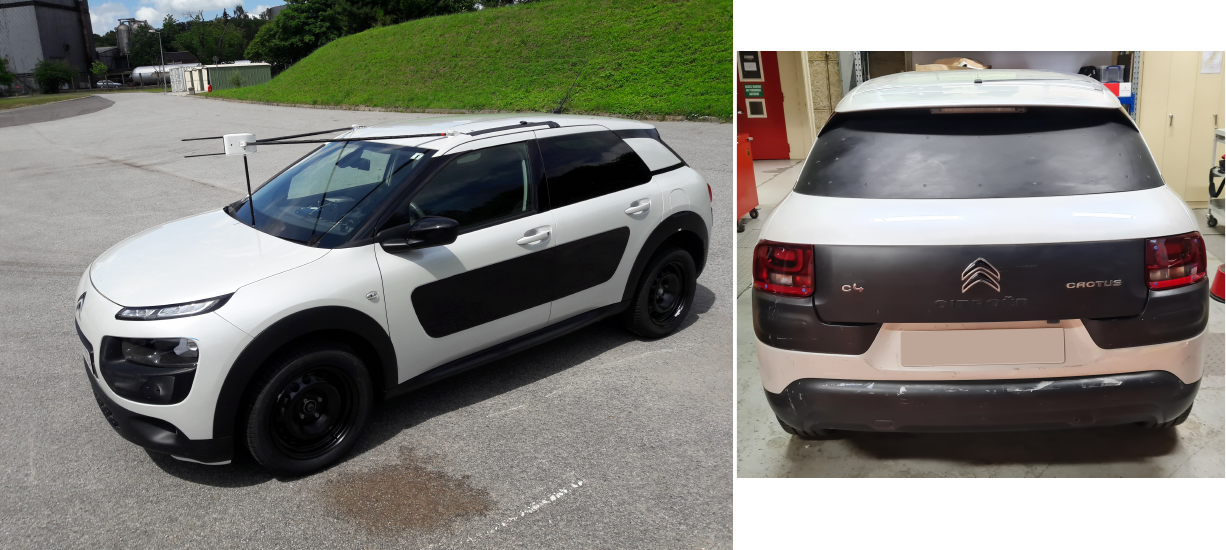}
    \caption{Front-side and back view of the Citroën C4 Cactus, respectively on the left and right-hand side.}
    \label{fig:C4_Cactus}
\end{figure}

\begin{figure}[ht!]
    \centering
    \includegraphics[width=0.44\textwidth]{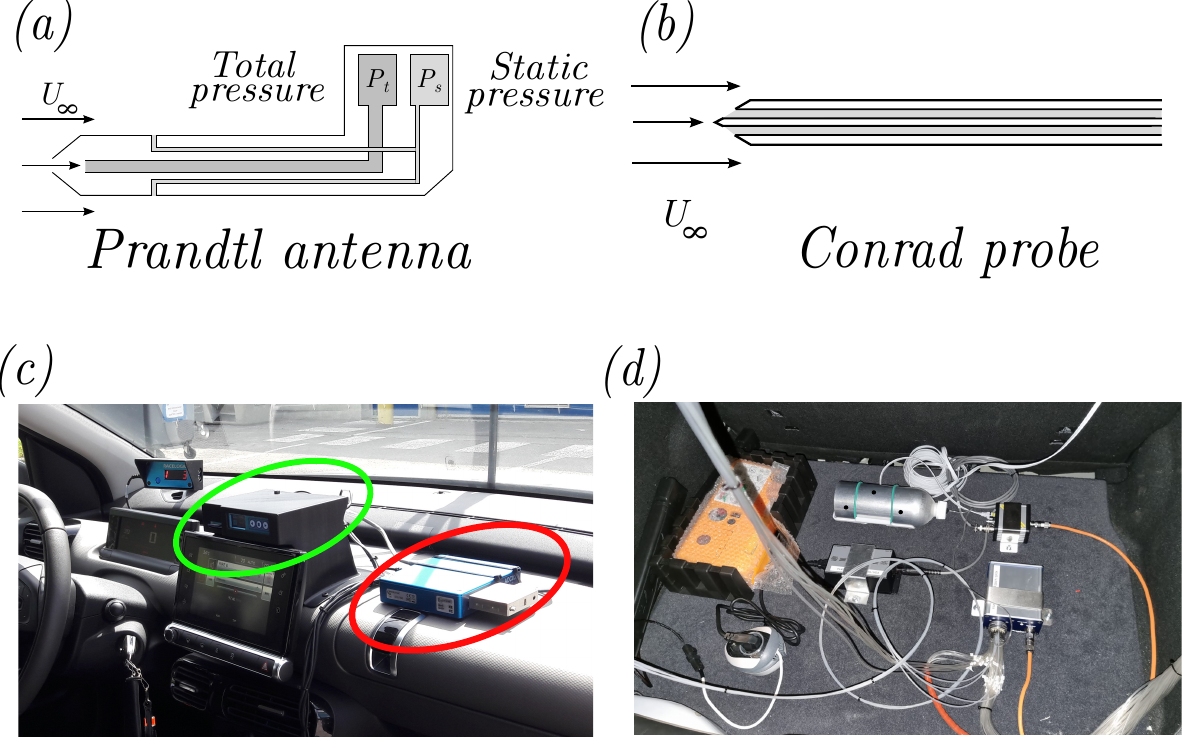}
    \caption{Measuring equipment installed on the vehicle. (a) Prandtl antenna schematisation. (b) Conrad probe schematisation. (c) GPS system mounted inside the vehicle. (d) Pressure scanner mounted in the trunk of the car.}
    \label{fig:outils_de_mesure}
\end{figure}

\begin{figure}[ht!]
    \centering
    \includegraphics[width=0.44\textwidth]{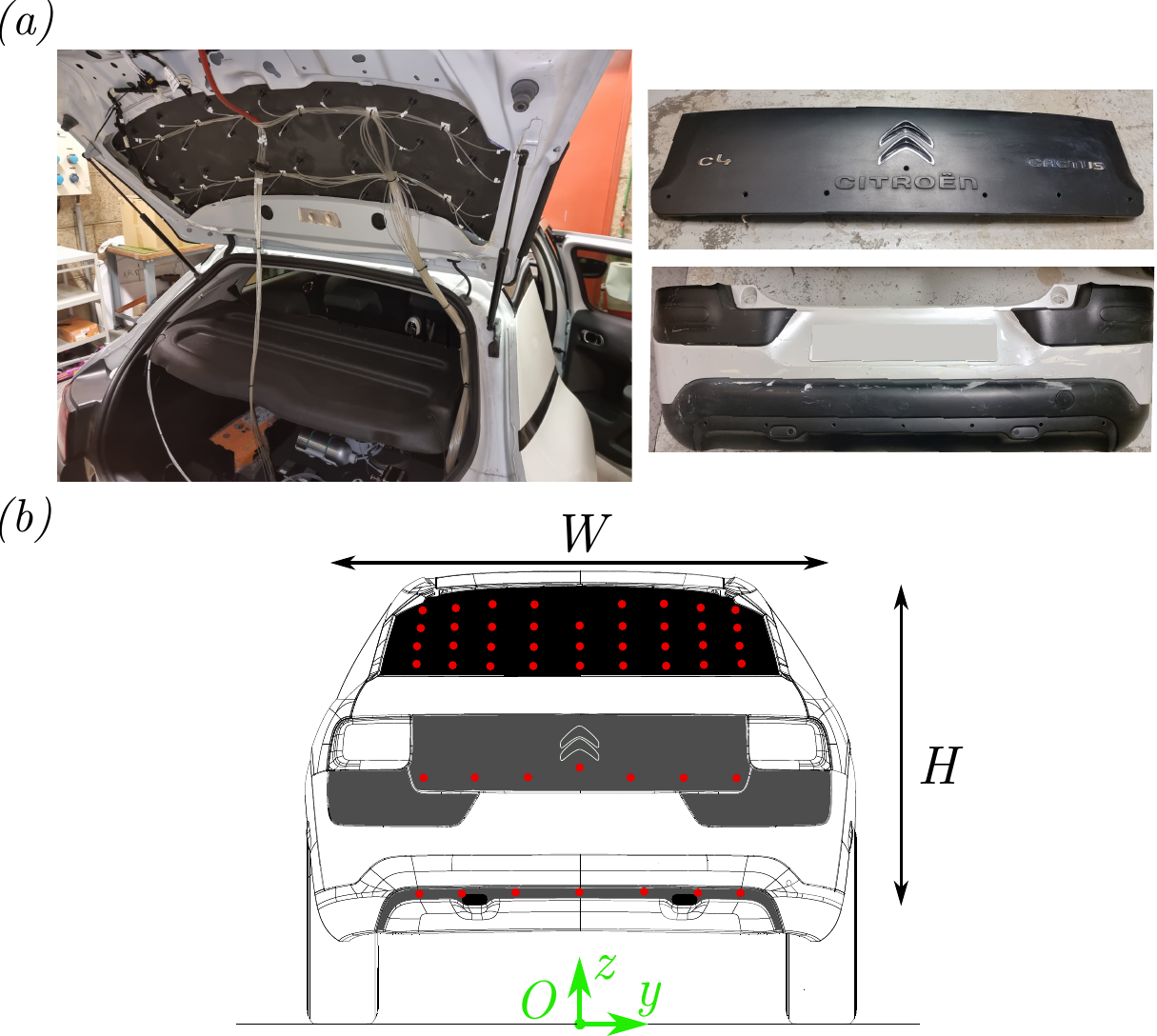}
    \caption{Pressure taps location on the vehicle's base. (a) Physical location of the pressure taps. (b) Position schematisation of the pressure taps.}
    \label{fig:prises_de_pression}
\end{figure}

A Citroën C4 Cactus, which is shown in Figure \ref{fig:C4_Cactus}, has been used with some added equipment. The vehicle has been equipped with a Prandtl antenna, a Conrad probe, a GPS system and 49 pressure taps at the base. All the equipment is shown in Figures \ref{fig:outils_de_mesure} and \ref{fig:prises_de_pression}.

Both Prandtl (Figure \ref{fig:outils_de_mesure} a) and Conrad (Figure \ref{fig:outils_de_mesure} b) probes have been constructed internally in a Stellantis' workshop, mainly following the instructions detailed in \citet{Chue1975}. They have been calibrated and tested by \citet{Thomann2018}. The Prandtl antenna grants an angular tolerance of $\pm 10^\circ$ while the Conrad probe has a maximum sensitivity of $60^\circ$ with a linear trend for $-15^\circ \leq \beta \leq 15^\circ$. The probes are respectively linked with a \emph{Texxis DPS 50 - 0.5} and a \emph{Texxis DPS 50 - 2.5} differential pressure scanners. In Figure \ref{fig:outils_de_mesure} (c) we can see the GPS system mounted inside the car. Highlighted in green the \emph{VBOX Racelogic} which is the GPS system itself while highlighted in red there is the \emph{CAN} box which gather the signals from the two probes and send them to the \emph{VBOX}. In Figure \ref{fig:outils_de_mesure} (d), the differential pressure scanner \emph{Scanivalve MPS4264}, mounted in the trunk of the car, is shown. It counts 64 total pressure entries and works within a range of 8" $H_2O$. Since it is a differential pressure scanner, this means that the pressure on each of the 64 taps is given by $p_i = {p_a}_i - p_r$, where ${p_a}_i$ is the absolute pressure on the $i_{th}$ tap, $p_r$ is the reference pressure, and $p_i$ is the measured value on the $i_{th}$ tap. In our specific case, the reference pressure is obtained using a pierced aluminum bottle mounted inside the trunk of the car. This allow to always have the static pressure in the trunk of the car while minimising the pressure fluctuations by acting like a low-pass filter. In Figure \ref{fig:prises_de_pression} (a) we can see the physical location of the pressure taps at the base of the car whereas in Figure \ref{fig:prises_de_pression} (b) a schematisation of the pressure tap position is given. In the latter, we can also visualize the location of the point considered as the origin as well as the width $W = 1150 \; mm$ and height $H = 900 \; mm$ of the base. The length of the vinyl tubes connected to the scanner varies considerably. At the base, this length can range from 0.7 meters to 2.6 meters, depending on the pressure tap considered. For each measurement channel and therefore each tube length, a phase and amplitude correction was carried out by convolution with a specifically determined transfer function (see \citet{Gravier2023}). To find out the transfer function, the method proposed by \citet{Tijdeman1965} was used taking into account the geometric properties of the circuit for each channel. In order to properly validate some sensitive information from the physical model derived by these authors, such as for example the radius of the vinyl tubes, a calibration was carried out on a specific test bench for several lengths. The calibration bench also made it possible to verify that, over the frequency range considered here (f $<\ 20Hz$) and after correction, the phase and module of the signals are accurately corrected.

The acquired pressure data is then post-processed to calculate the pressure coefficient:

\begin{equation}
C_{p_i} = \frac{p_i - p_r}{Q},
\label{eq:cp_defintion_E3}
\end{equation}

\noindent where $p_i$ represents the pressure measured at the $i_{th}$ pressure tap, $p_r$ is the reference pressure in the vehicle trunk, and $Q = \frac{1}{2} \rho_\infty U_\infty^2$ is the dynamic pressure, with $\rho_\infty$ as the air density and $U_\infty$ as the free-stream velocity. During on road tests, $U_\infty$ is the velocity deduced by the GPS system at the time of the measurement.

The following notation will be used for spatial ($\langle A\rangle$) and temporal ($\overline{A}$) averaging, respectively: 

$$\langle A\rangle (t_m) = \frac{1}{N}\sum_{n=1}^{N} A(\textbf{x}_n, \,t_m),$$

$$\overline{A} (x_n) = \frac{1}{M}\sum_{m=1}^{M} A(\textbf{x}_n,\,t_m),$$ 

\noindent where $N$ and $M$ are the number of pressure taps at the base and the number of time steps, respectively. 

However, it is necessary to note that measuring the reference pressure on-road remains a very complicated task because the slightest variation in the vehicle results in a variation in the pressure. Comparisons between the speed data obtained using the Prandtl antenna and those obtained using tachymetric measurements have shown that the reference pressure is not constant. The instantaneous difference between the pressure coefficient of each sensor and the spatial average of the pressure coefficient at the base at the same instant is defined as :

\begin{equation}
    C_p^*(\textbf{x}_n,\,t_m) = C_p(\textbf{x}_n,\,t_m) - \langle C_p (t_m) \rangle.
    \label{eq:Cp_star}
\end{equation}

Hereafter, we will always use $C_p^*(x_n,\,t_m)$ in order to eliminate any temporal variation in the reference pressure during a test. By using $C_p^*(x_n,\,t_m)$, we therefore focus on the asymmetry of the wake corresponding to the deviation from the spatial average of the instantaneous pressure distribution at the base.

\section{Experimental setup}
\begin{figure}[ht!]
    \centering
    \includegraphics[width=0.44\textwidth]{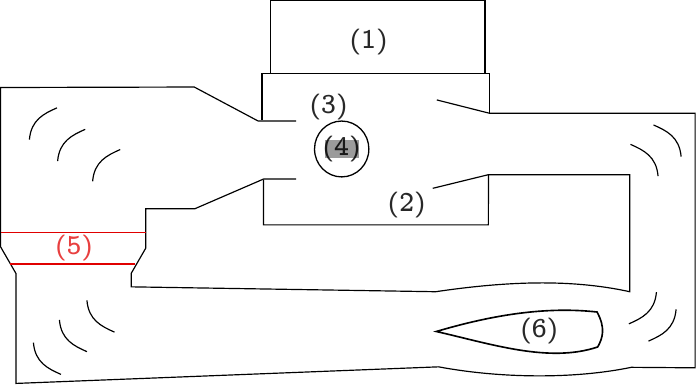}
    \caption{Schematisation of the Stellantis Wind tunnel in Orbassano (10043, Italy).}
    \label{fig:soufflerie}
\end{figure}

For this study, we were able to make use of the Stellantis wind tunnel located in Orbassano (10043, Italy) (\citet{Stellato2018}), the Stellantis track of La Ferté-Vidame (28340, France) and a RDE-like route (Yvelines, 78000, France), which are shown in the Figures \ref{fig:soufflerie} and \ref{fig:parcours_route} respectively. 

\subsection*{\textbf{Stellantis Wind Tunnel}}

\noindent From the control room (1), operators can activate the wind tunnel. The test section (2) is 3/4 open, with a variable nozzle section ranging from $22 \, \text{m}^2$ to $30.5 \, \text{m}^2$ in a plenum measuring 10.5 meters long, 12.2 meters wide, and 10.8 meters high. The turbulence rate at the center of the section is less than 0.5\%, and the boundary layer, at $X = 2.5 \, \text{m}$ from the exit of the convergent, has a thickness of $5 \, \text{mm}$ with the boundary layer suction function activated. The maximum test speed is $U_\infty = 215 \, \text{km/h}$ with a system of 5 moving belts, one under the car and the remaining four under the wheels to allow their rotation. The test speed for our tests is constant at $U_\infty = 110 \, \text{km/h}$. The car is positioned on a turntable (3) measuring 7 meters in diameter. Beneath the turntable, a six-component balance (4) is installed to measure aerodynamic forces. The vehicle is attached to the balance using four masts, with which we can also adjust the vehicle's ground clearance, set to the same value as measured with two people inside the vehicle for road tests. Through the heat exchanger (5), we control the temperature to remain constant at $T = 22^\circ \text{C}$ throughout the tests. The wind is set in motion with a motor-driven fan (6).

\subsection*{\textbf{On-road routes}}

\begin{figure}[ht!]
    \centering
    \includegraphics[width=0.44\textwidth]{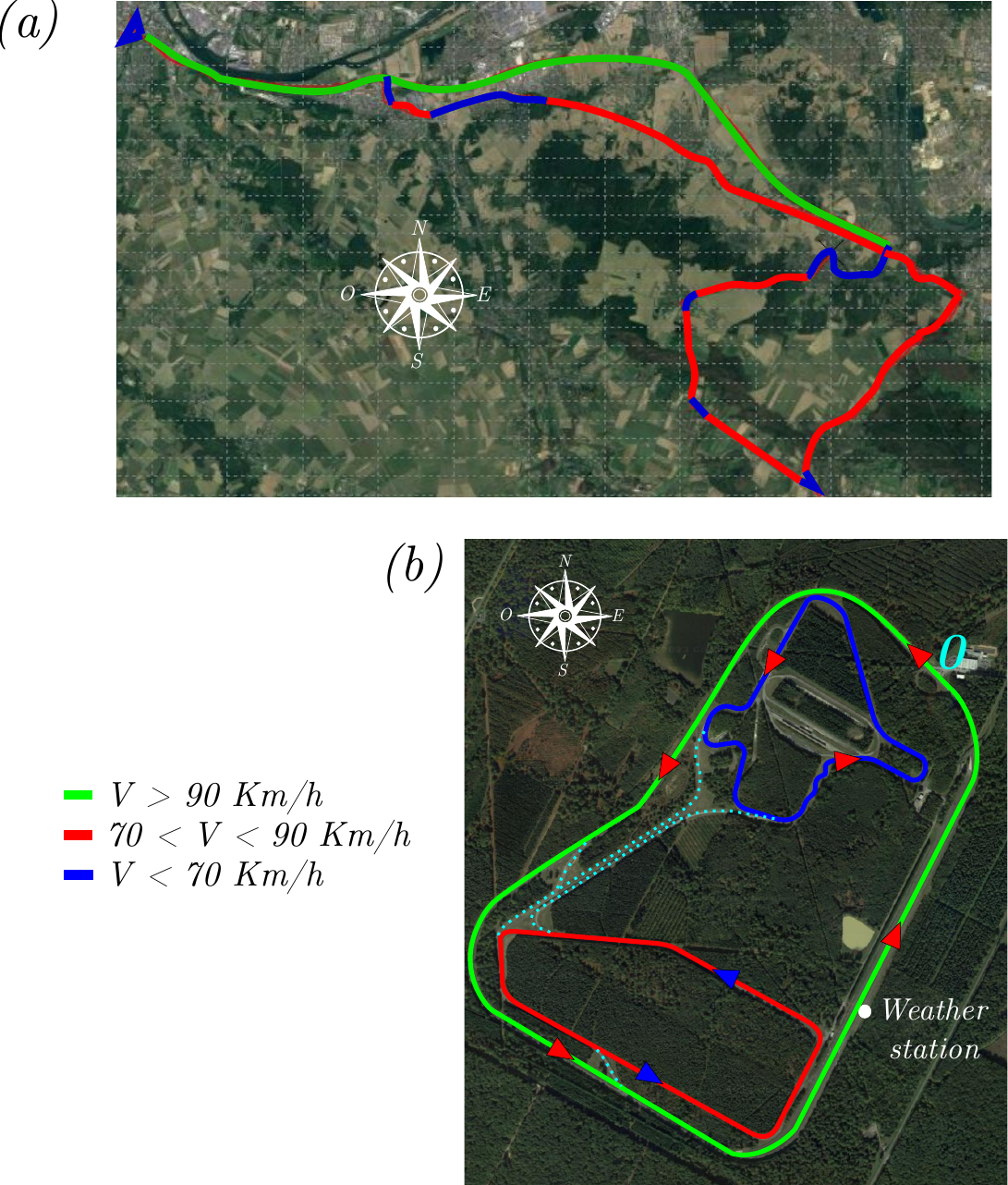}
    \caption{On-road routes visualisation. (a) RDE-like route in the Yvelines (78000, France). (b) Stellantis track of La Ferté-Vidame (28340, France).}
    \label{fig:parcours_route}
\end{figure}

\noindent The RDE-like route (part (a) of the Figure \ref{fig:parcours_route}) is a route that partially fullfils to the kilometer distribution in accordance with the new regulatory guidelines for the vehicle homologation process (WLTP). It measures approximately 68 kilometers and lasts about 1 hour. The whole route has, most of the time, allowed vehicle speed above $70 \; km/h$. This criterion was chosen because an internal study at Stellantis was conducted to determine from which speed aerodynamic efficiency becomes predominant compared to the rolling resistance of the vehicle (\citet{DaSilvaTuan2022}). This results in the following kilometer distribution: $\sim 43\%$ highway (in green), $\sim 46\%$ rural (in red), and $\sim 11\%$ urban (in blue). Atmospheric pressure information as well as wind speed and orientation were obtained from averages of several weather stations in the area.

Simultaneously, tests were conducted at the Stellantis site in La Ferté-Vidame (28340, France) (part (b) of the Figure \ref{fig:parcours_route}). This site consists of three tracks: the Outer Track (in green), the Bel Air Track (in red), and the Park Track (in blue). The colors correspond to the permitted speeds on the tracks, so the Outer Track corresponds to a highway route, the Bel Air Track corresponds to a rural route, and the Park Track corresponds to an urban route. We mainly conducted tests on the Outer Track and the Bel Air Track. The Outer Track measures approximately $\sim 6.5 \; km$ and includes four main turns. This track allows for a speed constantly over $90\; km/h$ throughout the test. The Bel Air Track measures approximately $\sim 3.3 \; km$ and includes several short-radius turns, allowing for a speed of $70 \leq V \leq 90$. These tracks are connected by connecting ramps (sky-blue dashed lines). The La Ferté-Vidame site also has a weather station capable of measuring wind speed and direction, as well as temperature and atmospheric pressure. All tests conducted on the tracks start at point 0 (in sky-blue), which corresponds to the main entrance of the tracks. The direction of travel is indicated by red triangles. However, each day, the direction of travel is reversed to be able to conduct tests in both directions and ensure that the cars and the track degrade uniformly and not biased by the direction of travel. In our tests, the number of people in the vehicle is always two. There is a driver and the person responsible for verifying the correct progress of the acquisition. The passenger sits in the rear seat where a PC mounting system has been installed to ensure safety requirements. A zero calibration is performed before each test to compensate for any pressure signal drift that may be present. This calibration is performed in a closed building to prevent small disturbances from affecting the process. Acquisition is initiated just before starting the test and stopped once the vehicle is stationary again.

\section{Wind tunnel testing}

The main objective of the wind tunnel sessions is to characterize the wake of the car in terms of global motions and to analyse the frequency of these motions. Subsequently, the same analysis will be carried out on the road to allow for a comparison of the results.

Proper Orthogonal Decomposition (POD) will be used in that respect. This analysis method was first introduced by \citet{Lumley1967} with the aim of decomposing the flow's turbulence into deterministic functions, each of which can capture a portion of the flow's turbulent kinetic energy. Several works have summarized this method and explained how to use it (\citet{Chatterjee2000, Cordier2008, Taira2017, Weiss2019}). 

In our specific case, the principle of POD, calculated here by the direct method after substracting the time average of $\overline{C^*_p(\textbf{x}_i,t)}$, consists of decomposing the vector $C^*_p(\textbf{x}_i,t)$ into a set of deterministic functions $\Phi^{(k)}(\textbf{x})$ that are modulated by N temporal coefficients $a^{(k)}(t)$ as follows:

\begin{equation}
C_p^* (\textbf{x}_i,t) =\overline{C^*_p(\textbf{x}_i,t)}+ \sum_{k=1}^{N_{modes}} a^{(k)}(t)\Phi^{(k)}(\textbf{x}_i).
\label{eq:POD_definition}
\end{equation}

\noindent It is important to recall that the functions $\Phi^{(k)}$ are orthogonal.

The decomposition is performed using singular value decomposition, that provides the functions $\Phi^{(k)}$, which we will call modes hereafter, as well as the eigenvector $\bm{\lambda}$, which contains the contribution of each mode, in decreasing order, to the total variance. In our study, the number of modes corresponds to the number of pressure sensors at the base, which is 49 modes in total. This means that we have a signal decomposition in the form:

\begin{equation*}
    C_p^*(\textbf{x}_i,t) = \overline{C^*_p(\textbf{x}_i,t)}+\sum_{k=1}^{49} a^{(k)}(t)\Phi^{(k)}(\textbf{x}_i),\\ 
\end{equation*}
\noindent with the total energy defined as follows:
\begin{equation*}
\xi = \sum_{i=1}^{N_{taps}} \overline{{C_{p_i}^*}^2} = \sum_{k = 1}^{N_{modes}} \lambda_{k}.
\end{equation*}

Additionally, thanks to the temporal coefficients of the modes ($a^{(k)}(t)$), it is possible to analyze the frequency characteristics of the projection of the velocity field onto each mode. 

In what follows, we will concentrate on the two first POD modes because they contain a significant contribution of the total variance. We remind that, by using $C_p^*(\textbf{x},\,t)$, we focus on the asymmetry of the wake.

\subsection*{\textbf{Static yaw angle analysis}}
 
\begin{figure}[ht!]
    \centering
    \includegraphics[width=0.46\textwidth]{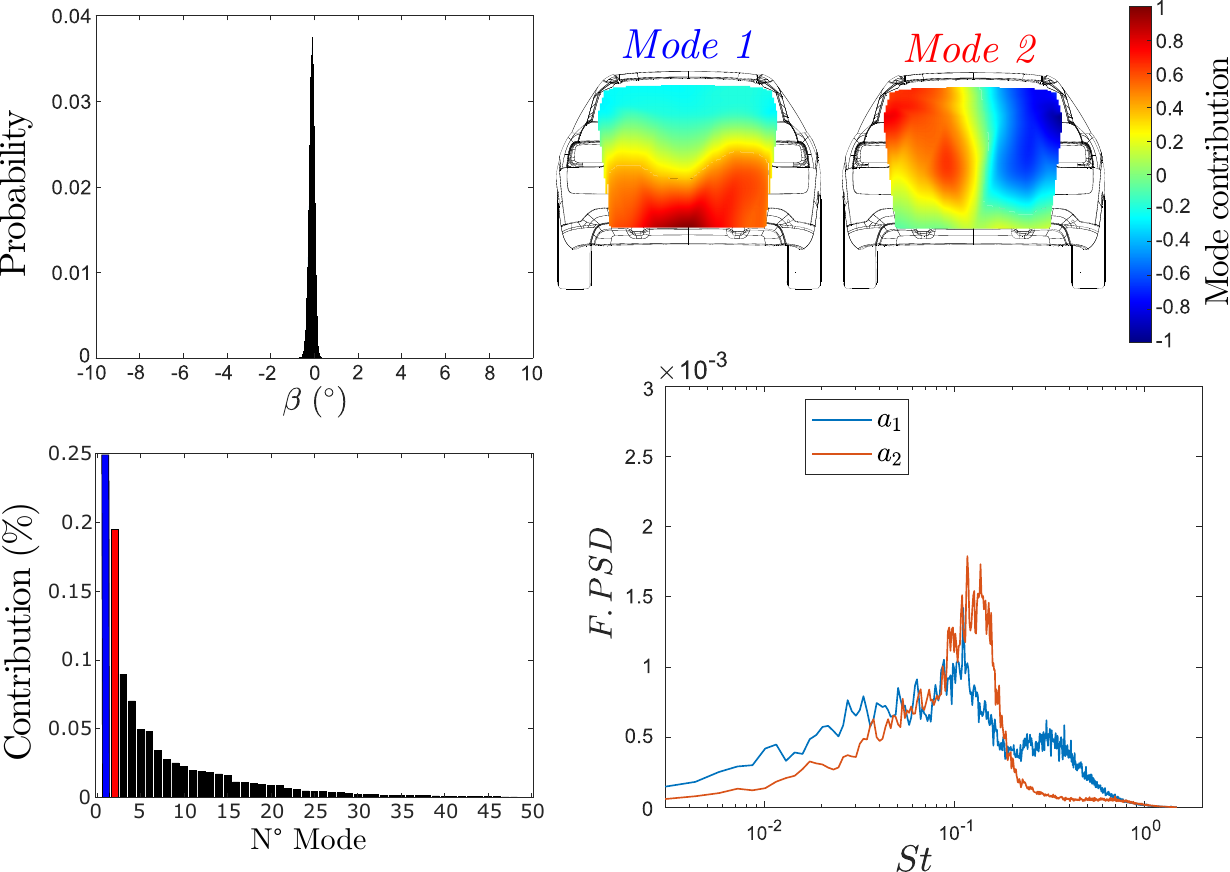}
    \caption{Results of the wake analysis for a fixed zero yaw angle in the wind tunnel. Presentation of the distribution of $\beta$, the first two modes, the percentage of their contribution to the total energy, as well as spectral analysis of the temporal coefficients associated with the two modes.}
    \label{fig:Beta_0_Soufflerie}
\end{figure}

The first analysis concerns a case with zero yaw angle ($\beta = 0^\circ$). The results are presented in Figure \ref{fig:Beta_0_Soufflerie}.

\noindent The distribution of $\beta$ shown in the Figure (top-left panel) has a mean $\mu_\beta = -0.3^\circ$ and a standard deviation $\sigma_\beta = 0.13^\circ$. It is not centered at $\beta = 0^\circ$ due to some slight human errors, such as alignment to the symmetry plane of the wind tunnel and alignment of the Conrad probe to the vehicle's symmetry plane. The latter error remains constant throughout our tests. This distribution of $\beta$ corresponds to typical cases analyzed by automotive manufacturers, i.e., a nearly constant yaw angle throughout the test. Under these conditions, we performed the POD analysis. The energy contribution of the 49 modes is presented in the bottom-left panel of the figure. We can observe that the first two modes contribute significantly to the total wake energy (approximately $45\%$), so we will analyze these two modes, which are presented in the top-right panel of the figure. It is interesting to note that these modes have well-defined footprints. For the first mode, we can associate this footprint with a vertical tilting of the wake. In contrast, the footprint of the second mode reveals a horizontal tilting. 
For this configuration, it is relevant to analyze the characteristic frequencies of these two asymmetries. To do this, we calculated the spectra of the two corresponding temporal coefficients with a frequency resolution $\Delta f = 5.10^{-2} \;Hz$. This frequency resolution corresponds to a division of the signal into 53 blocks, which will be the minimum number of blocks used subsequently. In the bottom-right panel of the figure, a pre-multiplied spectrum is presented showing the spectra of the two temporal coefficients. Regarding the frequencies, we chose to represent dimensionless frequencies to compare the different tests performed. For this purpose, we used the Strouhal number ($St = f.H/V$), which is the product of the frequency and the advective time scale ($H/V$). Note that the frequency range $f<20\;Hz$ discussed in section 2 corresponds here to a Strouhal number of $St <\ 0.6$, which means that phases and amplitudes are accurately corrected for all taps in the relevant frequency range. Moreover, from measurements performed with reduced scale models in a wind tunnel (\citet{Haffner2020a}) we do not expect significant higher frequency contributions for large scale events. We observe a large band distribution of energy for both $a^{(1)}$ and $a^{(2)}$, with a more defined peak at $St \approx 10^{-1}$ for $a^{(2)}$. As a measure of the important role of low frequencies we can observe that 62\% and 63\% of the energy of the mode for the first and second coefficients, respectively, are contained for Strouhal values less than $10^{-1}$.

\section{On-road results}
The response of the wake of the vehicle to on-road perturbations of the velocity seen by the vehicle is analysed in this section. We conducted numerous tests and a subset is shown in Table \ref{tab:Chapitre_2_Robustesse_essais}. The two on-road tests presented hereafter are the two test cases highlighted in bold in the Table. These two tests provide a comprehensive overview of the wake behavior under different conditions, allowing us to draw meaningful conclusions about the vehicle's wake behaviour when compared to wind tunnel tests at zero yaw angle. The robustness of these results is discussed in detail in \citet{Cembalo2024}. As shown in Table \ref{tab:Chapitre_2_Robustesse_essais}, the conclusions discussed below apply to all on-road tests with only minor differences on energy content in the POD decomposition or spectral energy density of the temporal coefficients. 

\begin{table*}[ht!]
\small
  \begin{center}
  \begin{tabular}{lcccccccc}
  Routes & $V_{wind} [km/h]$ & $O_{wind}$ & $P_{atm} [hPa]$ & $\mu_{\beta} \, [^\circ]$ & $\sigma_{\beta} \, [^\circ]$ & \% ($\Phi_1 + \Phi_2$)  & \% $a_1$/$a_2$ ($St \leq 10^{-1}$) & $\xi/\xi_0$\\ [3pt]\hline
\textbf{Track} & \textbf{22} & \textbf{North-West} & \textbf{975} & \textbf{-0.45} & \textbf{1.74} & \textbf{62} & \textbf{79 and 79} & \textbf{3.01}\\
Track & 25 & South-West & 975 & -0.19 & 1.68 & 62 & 79 and 79 & 5.52\\
Track & 13 & South & 974 & -0.44 & 1.52 & 63 & 78 and 79 & 3.57\\
Outer-Track & 14 & South & 979 & 0.02 & 1.37 & 56 & 76 and 81 & 2.71\\
Track & 16 & South-West & 978 & -0.08 & 1.54 & 63 & 77 and 79 & 2.79\\
Track & 8 & South & 987 & -0.31 & 1.04 & 69 & 74 and 69 & 3.01\\
Outer-Track & 3 & South-Est & 988 & -0.34 & 1.24 & 56 & 75 and 78 & 2.3\\
Outer-Track & 6 & Est & 987 & 0.01 & 1.15 & 58 & 77 and 77 & 2.43\\
Outer-Track & 6 & North-Est & 987 & 0 & 1.10 & 55 & 75 and 77 & 2.24\\
RDE-like & 8 & South-West & 1021 & -0.02 & 1.23 & 55 & 79 and 70 & 3.26\\
\textbf{RDE-like} & \textbf{9} & \textbf{South-Est} & \textbf{1021} & \textbf{-0.02} & \textbf{1.07} & \textbf{63} & \textbf{81 and 70} & \textbf{3.66}\\
RDE-like & 12 & North-Est & 1015 & -0.54 & 1.56 & 57 & 76 and 67 & 2.84\\
Highway & 12 & North-Est & 1015 & -0.18 & 1.63 & 53 & 72 and 72 & 1.98\\
RDE-like & 21 & North & 1009 & -0.18 & 1.65 & 55 & 76 and 72 & 3.66\\
RDE-like & 25 & North-Est & 1009 & 0.1 & 1.4 & 55 & 76 and 70 & 3.03\\
RDE-like & 3 & South-West & 1022 & 0.02 & 1.41 & 63 & 72 and 65 & 6.22\\
RDE-like & 6 & West & 1021 & 0.1 & 1.54 & 60 & 69 and 69 & 2.85\\[3pt]\hline
  \end{tabular}
  \caption{Some statistics to show the robustness of the on-road tests. Hereafter, the two tests in bold will be analysed in detail.}
  \label{tab:Chapitre_2_Robustesse_essais}
  \end{center}
\end{table*}

\begin{figure}[ht!]
    \centering
    \includegraphics[width=0.46\textwidth]{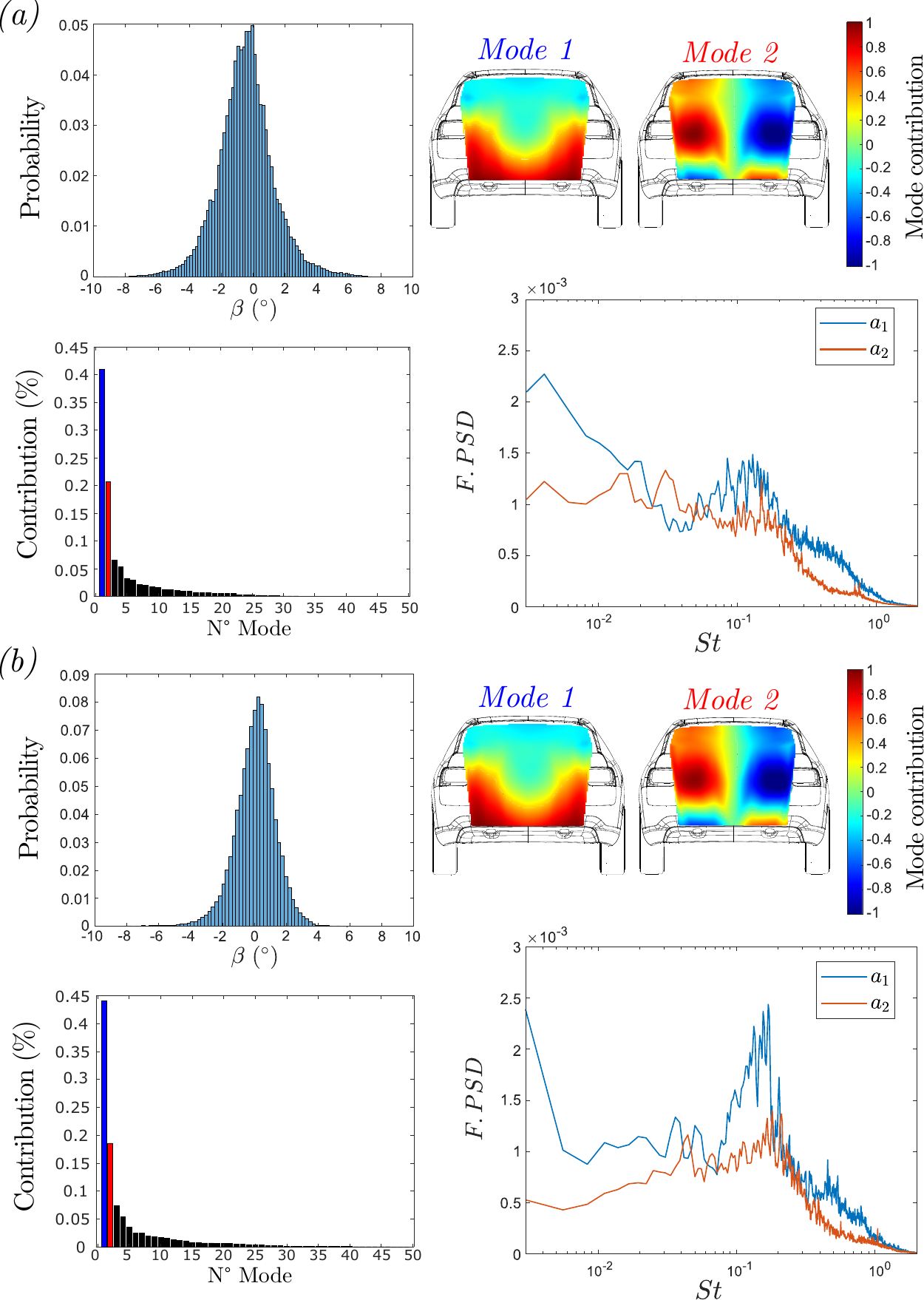}
    \caption{Results of wake analysis on track and on the RDE-like route. Presentation of the distribution of $\beta$, the first two modes, and the percentage contribution to the total energy as well as spectral analysis of the temporal coefficients associated with the two modes. (a) Results for test 08 on August 6, 2021, at the La Ferté-Vidame track. (b) Results for test 04 on July 22, 2021, on the RDE-like route.}
    \label{fig:Beta_circuit_route_resultat}
\end{figure}

In Figure \ref{fig:Beta_circuit_route_resultat} (a), the results are from a test on the La Ferté-Vidame track, while in part (b), it corresponds to a test on the RDE-like route. Atmospheric pressure information as well as wind speed and orientation for the two tests are detailed in Table \ref{tab:Chapitre_2_Robustesse_essais}.


Regarding the test conducted at La Ferté-Vidame, the direction of travel was counterclockwise, as indicated by the red triangles in Figure \ref{fig:parcours_route}. The test route includes approximately 4 laps on the Outer track and 2 laps on the Bel Air track. There is very little traffic on this road closed to public. When compared to RDE test conditions, the surroundings of the road on the Stellantis track is repeated at each lap while it displays more diversity on public roads. 

It is interesting to note that the distributions of $\beta$ are very different to that observed in Figure \ref{fig:Beta_0_Soufflerie}. On the track (part (a) of the Figure \ref{fig:Beta_circuit_route_resultat}), the distribution has a mean $\mu_\beta = -0.45^\circ$ and a standard deviation $\sigma_\beta = 1.74^\circ$, while on the RDE-like route (part (b) of the Figure \ref{fig:Beta_circuit_route_resultat}), the distribution has a mean $\mu_\beta = -0.02^\circ$ and a standard deviation $\sigma_\beta = 1.07^\circ$. On the track, the distribution is not centered at $\beta = 0^\circ$, which is consistent with the counterclockwise direction of the laps. Additionally, the wind speed was higher during the track test, which may explain the difference observed in the standard deviation.

POD modes 1 and 2 have a very clear large scale signature corresponding to respectively vertical and horizontal balance of the wake. They are very similar for both on-road tests. They are also qualitatively similar to the main POD modes obtained in the wind tunnel at zero yaw angle. However, some details differ, for example the signature of mode 2 at the bottom part of the base. The contribution of these two modes to the total energy is now 63\%. This is a very significant increase when compared to the wind tunnel tests (45\% contribution for these two modes). 

On the RDE-like route, a slight increase in the energy contained in mode 1 (about 44\%) and a slight decrease in mode 2 (about 19\%) is observed. However, the sum of the energy contained in the two modes remains almost unchanged. Looking now at the premultiplied spectra of the temporal coefficients associated to modes 1 and 2, an increase of the energy in the low frequency domain is observed for the first mode. For all these tests, 70\% to 80\% of the energy content is measured for Strouhal numbers lower than 0.1. Note that, for on-road situations, the vehicle velocity $V$ used in the computation of the Strouhal number in figure \ref{fig:Beta_circuit_route_resultat} is the average vehicle velocity during the test. 

\begin{figure}[ht!]
    \centering
    \includegraphics[width=0.46\textwidth]{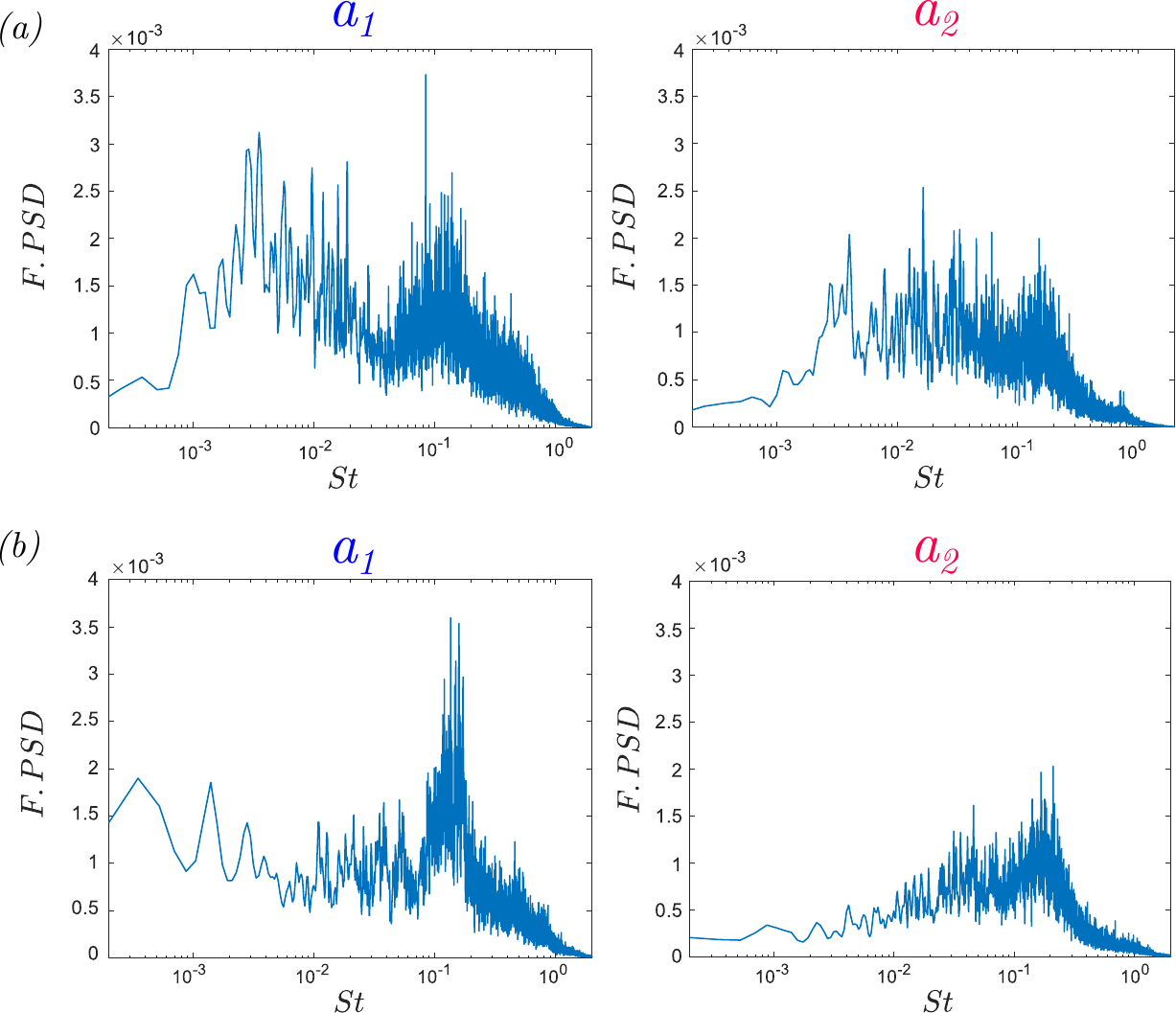}
    \caption{Results of the spectral analysis of the temporal coefficients with a frequency resolution of $\Delta f = 5.10^{-3} [Hz]$ for the track and RDE-like route tests. (a) Coefficients $a^{(1)}$ and $a^{(2)}$ on the track. (b) Coefficients $a^{(1)}$ and $a^{(2)}$ on the RDE-like route.}
    \label{fig:coeff_temps_resolution_freq}
\end{figure}

A higher frequency resolution ($\Delta f = 5.10^{-3} \;Hz$) was obtained by using longer blocks and therefore a lower number of blocks for averaging the spectra ($N_{blocks} \leq 13$). The results are presented in Figure \ref{fig:coeff_temps_resolution_freq}. The presence of a large band contribution for Strouhal numbers lower than  $10^{-2}$ is very clear, especially for the first mode corresponding to the vertical wake balance (Figure \ref{fig:Beta_circuit_route_resultat}). We observe that very low frequencies (down to $St \approx 10^{-3}$) contribute to the power spectrum. At the average speed maintained during the track test, the distance traveled by the car during a period $T = 1/f$ can be estimated as $d\approx V.T = H/St$ which is very large for Strouhal numbers lower than $10^{-2}$. We stress that a quasi steady response of the wake is expected for these low Strouhal numbers. This is an important aspect because these low frequencies do not only come from wind turbulence. They probably also originate from the variation in the flow seen by the car due to a variable environment having a wide range of longitudinal scales along the road. Indeed, driving along road portions having similar characteristics (freely exposed to side winds, protected in a forest, exposed to urban area features, ...) will induce low frequency components corresponding to the time taken to travel along these portions. Of course, sharp changes in wind directions occur while entering to - or emerging from - a sheltered portion. However, careful experimental study by \citet{Haffner2020a} using a model geometry at reduced scale shows that natural or provoked wake switching between right/left asymmetry is a slow process taking place in approximately 30 convective time scales. This corresponds to $St\approx 1/30$. The slow reorganisation of the large scale wake therefore acts as a low-pass filter that cuts high frequency disturbances.

\section{Statistical analysis at varying yaw angle in the wind tunnel}

\begin{figure}[ht!]
    \centering
    \includegraphics[width=0.46\textwidth]{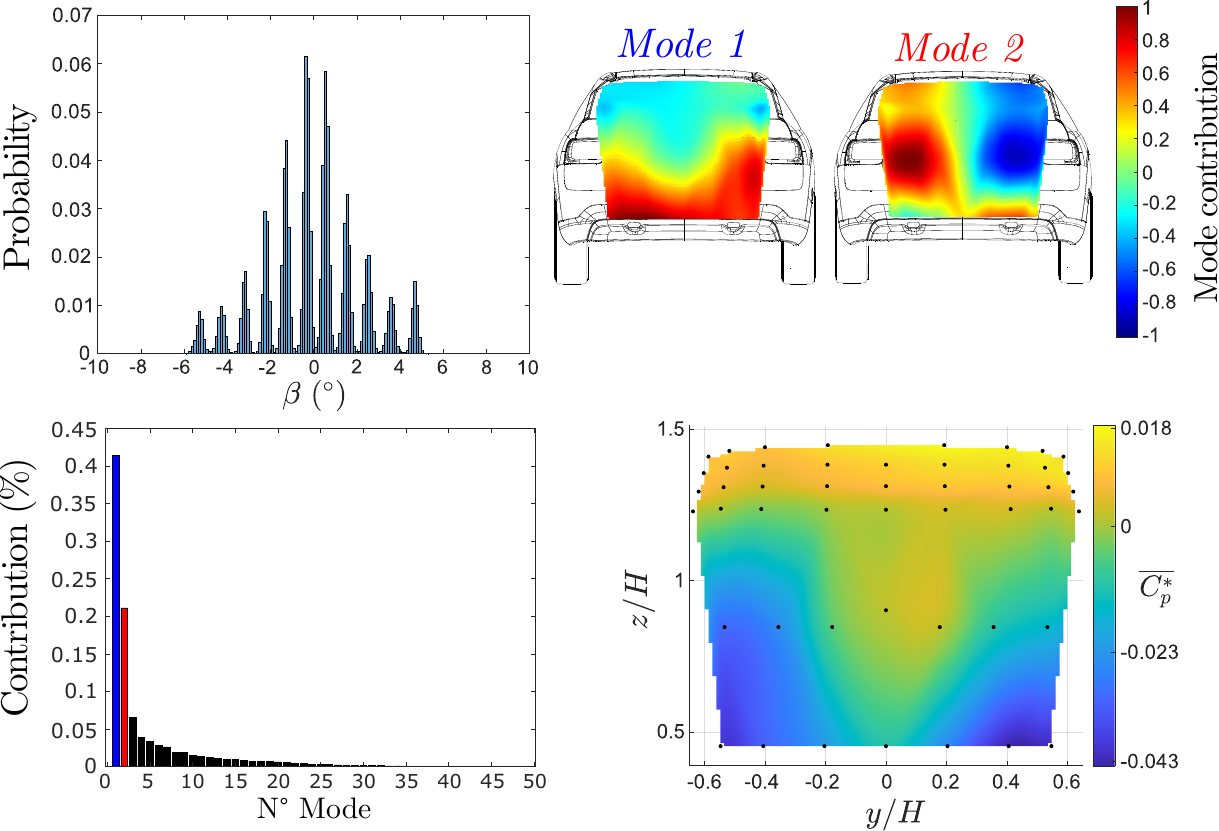}
    \caption{Results of the wake analysis for a varying yaw angle in the wind tunnel. Presentation of the distribution of $\beta$, the first two modes and the percentage of their contribution to the total energy. The time averaged value $\overline{C^*_p(\textbf{x},t)} $ at the base for the full test is shown on bottom right plot}
    \label{fig:Beta_dynamique_Soufflerie}
\end{figure}

Going back to the wind tunnel, we can try to reproduce the yaw angle distribution observed on the road. A subset of yaw angles of the vehicle is first chosen. One single continuous wind tunnel test of approximately 20 minutes - the total number of sample is 120687 at a sampling rate of 100 Hz - is used for the test shown in Figure \ref{fig:Beta_dynamique_Soufflerie}. During the test, our choice is to impose a variable measurement time for each vehicle yaw angle of the subset in order to have a data set representative of the statistical distribution of yaw angles observed on the road.

Modes 1 and 2 are very similar to on road modes. We observe that nearly 63\% of the turbulent kinetic energy is contained in the first two modes. This is also very similar to the energy distribution for on-road tests. Moreover, the energy contained in the first mode has almost doubled compared to the static yaw angle test, which seems to be due to the variation of the yaw angle along the tests, similarly to what we could observe on the road. This may be a good indication of the interest for making these quasi steady tests during vehicle development to test the robustness of a vehicle design to yaw angle perturbations. 

\begin{figure}[ht!]
    \centering
    \includegraphics[width=0.46\textwidth]{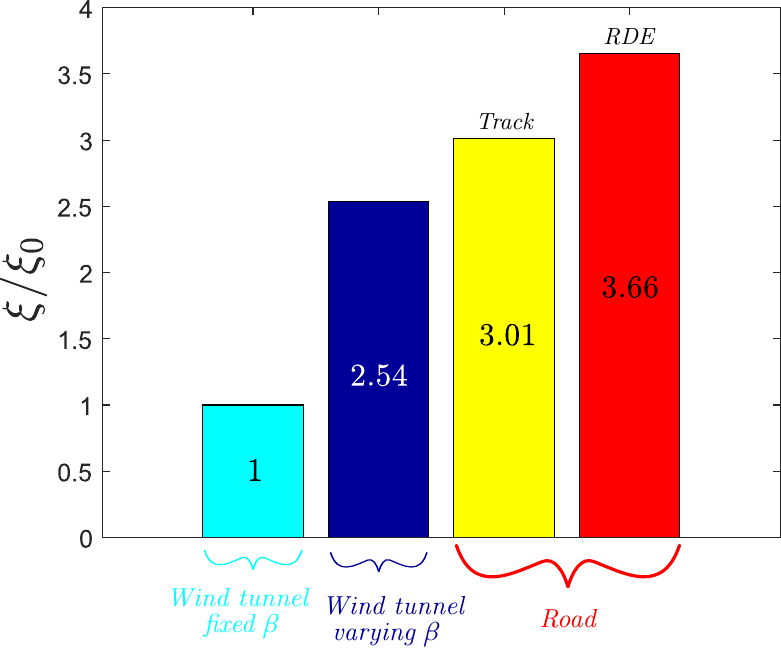}
    \caption{Comparison of total variance $\xi$ for the different analyzed tests. In sky blue, the wind tunnel case with fixed $\beta$. In dark blue, the wind tunnel case with varying $\beta$. In yellow and red, the cases on track and on the RDE-like route, respectively on the left and on the right.}
    \label{fig:Comparaisons_energies}
\end{figure}

The total energy $\xi$ measured in the different tests is compared in Figure \ref{fig:Comparaisons_energies} to that measured in the wind tunnel for a fixed yaw angle. This ratio increases significantly for the test with dynamic variation of $\beta$ in the wind tunnel, reaching more than double the reference value ($\xi / \xi_0 \approx 2.5$). The total energy further increases on the road. On the track, the difference is similar to the cases with dynamically varying $\beta$ in the wind tunnel ($\xi / \xi_0 \approx 3.01$), while on the RDE-like route, the total variance increases even more ($\xi / \xi_0 \approx 3.66$) compared to the value observed in the wind tunnel for fixed yaw, which is a very significant difference.

The imposed quasi static variations of the yaw angle are interesting in order to widen the statistical distribution of the projection of instantaneous pressure footprints on the main POD modes that represent the wake vertical and horizontal asymmetries. These modes are very similar to on-road main modes and this is an interesting property for wind tunnel tests to be confirmed on other vehicle geometries.

\begin{figure}[ht!]
    \centering
    \includegraphics[width=0.46\textwidth]{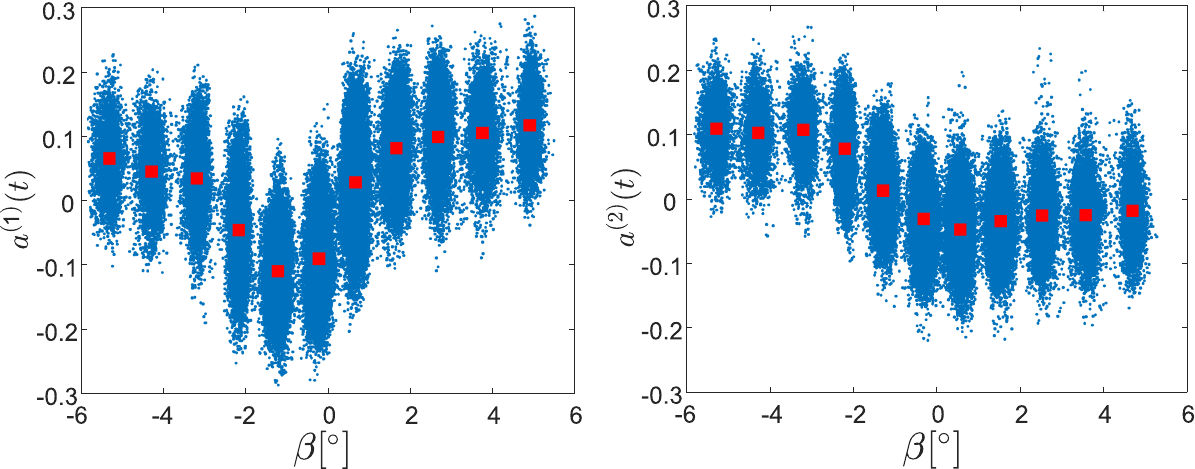}
    \caption{Distribution of the temporal coefficients $a^{(1)}$ and $a^{(2)}$ as function of the yaw angle $\beta$ in the varying yaw angle test. The red dot is the conditional average of the temporal coefficient for each orientation of the vehicle.}
    \label{fig:plot_A1_beta}
\end{figure}

Figure \ref{fig:plot_A1_beta} shows the distribution of the first and second temporal coefficients as function of the yaw angle. Each group corresponds to a prescribed orientation of the vehicle during the dynamic tests. The red dot is the conditional average of the temporal coefficient for each orientation of the vehicle. We see that the POD decomposition is able to describe a large span of wake states at the base. It is well known that drag increases with yaw angle. For the largest positive or negative yaw angles considered here, the change of sign of $a^{(2)}$ corresponds to a change in left/right asymmetry. As $a^{(1)}$ is concerned, we see that positive values correspond to large yaw angles while small yaw angles - and therefore minimum drag - are associated statistically to negative values of $a^{(1)}$.
The time averaged value $\overline{C^*_p(\textbf{x},t)}$ at the base for the full test is also shown in Figure \ref{fig:Beta_dynamique_Soufflerie}. This mean distribution displays approximately a left/right symmetry and a positive pressure gradient (pressure increases when moving upward). Using the POD decomposition a low order representation of $C^*_p(\textbf{x},t) $ can be obtained using the first two modes only. From the signature of the mean distribution, from the spatial distribution of the two dominant modes shown in Figure \ref{fig:Beta_dynamique_Soufflerie} and from the conditional values of the temporal coefficients discussed above, it is easy to figure out that the wake is evolving from a positive vertical asymmetry at about zero yaw to a horizontal asymmetry at larger yaw. Indeed, at larger yaw, a positive value of $a^{(1)}$ reduces the vertical gradient while a non zero $a^{(2)}$ corresponds to horizontal pressure gradients. Considering the similarities with on-road results, we therefore conclude that a low frequency exploration of such wake large scale states is performed during on-road travel. Therefore, even a slow closed loop control able to maintain a prescribed wake balance would be interesting for drag reduction. This approach is proposed for a reduced scale model using actuated flaps located at the back of a model vehicle in \citet{Cembalo2024} and \citet{Cembalo2024a}.

\section{Key findings and Conclusions}
The aim of this research work was to analyse the effect of on-road perturbations on the asymmetry of a vehicle wake. We described the vehicle used for testing in a full-scale wind tunnel, on a track and on a RDE-like route. Additionally, we detailed the measurement systems and experimental setup. 
To analyze the wake characteristics in the wind tunnel and on the road, Proper Orthogonal Decomposition (POD) of the pressure distribution on the base of the vehicle was employed. The main results are summarized below:

\begin{itemize}
\item Wind tunnel tests with fixed yaw substantially deviate from road conditions, showing lower modal energy content and different wake modal footprints, yet still corresponding to similar large scale vertical or horizontal wake asymmetries. The low-frequency energy content of the temporal coefficients of the mode is significantly different from that observed on-road.

\item Road tests exhibit a nearly normal distribution of yaw angle, with values consistently between $\pm 10^\circ$ and $-5^\circ \leq \beta \leq 5 ^\circ$ for over 95\% of the time. POD analysis also reveals two primary modes, associated with vertical and horizontal wake asymmetries, respectively. More than 60\% of the total energy is carried by these two modes. Low frequencies (even very low ones down to $St \approx 10^{-3}$) play a major role, corresponding to a quasi-static perturbation domain. Indeed, over 70\% of the energy of the two main modes is contained within frequencies $St \leq 10^{-1}$;

\item A wind tunnel run can be designed by selecting a subset of vehicle yaw and a variable measurement time for each yaw. The large data set obtained is then representative of the statistical distribution of yaw angles observed on the road. This single run captures phenomena similar to those observed on the road. The footprint of the two main modes are very similar. The energy content in the first two modes is higher and closer to that measured on road. These tests also show that multiple large scale wake states contribute to the first two modes. Considering the similarities with on-road results, we conclude that a low frequency large span exploration of such wake states is performed during on-road travel.
\end{itemize}

Further studies should complete this analysis. It would be interesting to analyse the effect of pitch angle variations on these large scale modes.  Moreover, the tests have been conducted on a single vehicle. It would be interesting to see the robustness of these results on other types of vehicle. 
We also stress that a quasi steady response of the wake is expected for the energy containing low Strouhal numbers observed on the road. Therefore, the use of steady CFD computations for a subset of yaw angles is interesting to evaluate a vehicle on-road characteristics. Moreover, even a slow closed loop control able to maintain a prescribed wake balance would be promising for drag reduction strategies. This is an ongoing work on 2/5 reduced scale models using recursive subspace-based predictive control as proposed in \citet{Cembalo2024} and \citet{Cembalo2024a}.
\section*{Acknowledgements}

The authors would like to warmly thank Jean Charles Boueilh for invaluable support during the experiments, as well as Yann Goraguer for moral and technical assistance.

\section*{Funding}

This research project was funded by STELLANTIS and Ministry for Higher Education and Research. Agostino Cembalo wishes to acknowledge support from ANRT scholarship.

\section*{Competing interest}

The authors report no conflict of interest.

\bibliography{article}

\end{document}